\documentclass[aps,prl,10pt,twocolumn, superscriptaddress, nofootinbib]{revtex4}
\usepackage{mathrsfs, amssymb, amsmath}  
\usepackage{epsfig, cancel}
\usepackage{latexsym}
\usepackage{natbib, comment}
\usepackage{url}
\usepackage{dcolumn}
\usepackage{multirow}
\usepackage{color}
\usepackage{cancel}
\usepackage{soul}
\usepackage[normalem]{ulem}
\usepackage{amsfonts,amssymb,amsmath, txfonts}
\usepackage{graphicx,epsfig}
\usepackage{psfrag}
\usepackage{hyperref}
\hypersetup{colorlinks=true}
\usepackage{mathtools}
\usepackage{enumitem}
\usepackage{float}
\usepackage[dvipsnames]{xcolor}
\usepackage{xcolor}
\hypersetup{ linktoc=all,
    colorlinks, linkcolor={brightpink},
    citecolor={blue}, urlcolor={blue}
}
\definecolor{rosy}{RGB}{230,235,252}
\definecolor{myframetitle}{RGB}{90,89,170}
\definecolor{myblocktitle}{RGB}{140,185,249}
\definecolor{mytitle}{RGB}{10,80,26}

\definecolor{darkgreen}{RGB}{27,130,45}
\definecolor{darkblue}{rgb}{0,0,0.3}
\definecolor{darkred}{rgb}{0.7,0,0}

\definecolor{light gray}{RGB}{220,220,220}
\definecolor{dark purple}{RGB}{108,0,217}
\definecolor{pink}{RGB}{190,20,100}
\definecolor{orang}{RGB}{193,63,0}
\definecolor{green}{RGB}{11,98,17}
\definecolor{darkpink}{RGB}{153,0,76}
\definecolor{bluegreen}{RGB}{0,102,102}
\definecolor{greenlagan}{RGB}{0,102,0}
\definecolor{redgreen}{RGB}{102,102,0}
\definecolor{Redgreen}{RGB}{153,76,0}
\definecolor{vividviolet}{rgb}{0.62, 0.0, 1.0}
\definecolor{amaranth}{rgb}{0.9, 0.17, 0.31}
\definecolor{palatinateblue}{rgb}{0.15, 0.23, 0.89}
\definecolor{brightpink}{rgb}{1.0, 0.0, 0.5}
\definecolor{cornflowerblue}{rgb}{0.39, 0.58, 0.93}
\definecolor{deepcarminepink}{rgb}{0.94, 0.19, 0.22}
\definecolor{radicalred}{rgb}{1.0, 0.21, 0.37}
\def\H0{{\text{H}\hspace*{-2.05mm}\text{H} 0\hspace*{-1.35mm}0\ }}

\def\be{\begin{equation}}
\def\ee{\end{equation}}
\def\beq{\begin{equation}}
\def\eeq{\end{equation}}
\def\bea{\begin{eqnarray}}
\def\eea{\end{eqnarray}}
\newcommand{\dd}{\textrm{d}}

\begin{document}

\title{Revealing Intrinsic Flat $\Lambda$CDM Biases with Standardizable Candles}

\author{E. \'O Colg\'ain} 
%\email{eoin@sogang.ac.kr}
\affiliation{Center for Quantum Spacetime \& Dept. of Physics, Sogang University, Seoul 121-742, Korea}
\author{M.M. Sheikh-Jabbari}
\affiliation{School of Physics, Institute for Research in Fundamental Sciences (IPM), P.O.Box 19395-5531, Tehran, Iran}
\author{R. Solomon}
\affiliation{HEPCOS, Department  of  Physics,  SUNY  at  Buffalo,  Buffalo,  NY  14260-1500, USA}
\author{G. Bargiacchi}
\affiliation{Scuola Superiore Meridionale,
                Largo S. Marcellino 10,
                80138, Napoli, Italy}
\affiliation{Istituto Nazionale di Fisica Nucleare (INFN), Sez. di Napoli,
                Complesso Univ. Monte S. Angelo, Via Cinthia 9,
                80126, Napoli, Italy}
\author{S. Capozziello}
\affiliation{Dipartimento di Fisica "E. Pancini" , Universit\'a degli Studi di  Napoli "Federico II"\\
                Complesso Univ. Monte S. Angelo, Via Cinthia 9
                80126, Napoli, Italy}
\affiliation{Scuola Superiore Meridionale, 
                Largo S. Marcellino 10,
                80138, Napoli, Italy}
\affiliation{Istituto Nazionale di Fisica Nucleare (INFN), Sez. di Napoli,
                Complesso Univ. Monte S. Angelo, Via Cinthia 9,
                80126, Napoli, Italy}
\author{M. G. Dainotti}
\affiliation{National Astronomical Observatory of Japan, 2 Chome-21-1 Osawa, Mitaka, Tokyo 181-8588, Japan}
\affiliation{The Graduate University for Advanced Studies, SOKENDAI, Shonankokusaimura, Hayama, Miura District, Kanagawa 240-0193, Japan}
\affiliation{Space Science Institute, Boulder, CO 80301, USA}

\author{D. Stojkovic}
\affiliation{HEPCOS, Department  of  Physics,  SUNY  at  Buffalo,  Buffalo,  NY  14260-1500, USA}

\begin{abstract}
Emerging high redshift cosmological probes, in particular quasars (QSOs), show a preference for larger matter densities, $\Omega_{m} \approx 1$, within the flat $\Lambda$CDM framework. Here, using the Risaliti-Lusso relation for standardizable QSOs, we demonstrate that the QSOs recover the \textit{same} Planck-$\Lambda$CDM Universe as Type Ia supernovae (SN), $\Omega_m \approx 0.3$ at lower redshifts $ 0 < z \lesssim 0.7$, before transitioning to an Einstein-de Sitter Universe ($\Omega_m =1$) at higher redshifts $z \gtrsim 1$. We illustrate the same trend, namely increasing $\Omega_{m}$ and decreasing $H_0$ with redshift, in SN but poor statistics prevent a definitive statement. We explain physically why the trend may be expected and show the intrinsic bias through non-Gaussian tails with mock SN data. Our results highlight an intrinsic bias in the flat $\Lambda$CDM Universe, whereby $\Omega_m$ increases, $H_0$ decreases and $S_8$ increases with effective redshift, thus providing a new perspective on $\Lambda$CDM tensions; even in a Planck-$\Lambda$CDM Universe the current tensions may be expected.  
\end{abstract}

\maketitle

\section{Introduction}
Our current understanding of the Universe, as described by the flat $\Lambda$CDM model, largely rests upon three pillars:  Type Ia supernovae (SN) cosmology \cite{Riess:1998cb, Perlmutter:1998np}, the cosmic microwave background (CMB) \cite{Planck:2018vyg} and baryon acoustic oscillations (BAO) \cite{SDSS:2005xqv}. While these observables show perfect agreement on $\Omega_{m} \approx 0.3$, recent direct checks of the flat $\Lambda$CDM model in the late Universe challenge the current paradigm \cite{Verde:2019ivm, DiValentino:2021izs, Abdalla:2022yfr}. The crux of this letter is that cosmological probes already point to evolution of matter density $\Omega_m$, and consequently $H_0$, within the flat $\Lambda$CDM model. 

 Risaliti \& Lusso have  introduced a relation between X-ray  and UV QSO luminosities, respectively $L_{X}, L_{UV}$, for cosmological purposes \cite{Risaliti:2015zla, Risaliti:2018reu}: 
\be
\label{RL}
\log_{10} L_{X} = \beta + \gamma \log_{10} L_{UV},  
\ee
where $\beta$ and $\gamma$ are fitting constants. This relation follows from an empirical relation between the corresponding fluxes and  it has been shown that it is robust to selection biases and redshift evolution  \cite{Dainotti:2022rfz}, so the relation appears intrinsic to QSOs. It has been shown that the slope $\gamma \approx 0.6$ is robust across luminosities and redshifts \cite{Vignali:2002ct, Just:2007se, Lusso:2009nq, Salvestrini:2019thn, Bisogni:2021hue}. In contrast to other QSO standardization methods \cite{Watson:2011um, Wang:2013ha, LaFranca:2014eba,Solomon:2021jml}, equation (\ref{RL}) represents an approach that is extremely powerful, as it can be applied across extended redshifts and luminosities. 

Here, we largely highlight synergies between QSOs and SN within flat $\Lambda$CDM. First, we show that the Risaliti-Lusso QSOs recover Planck-$\Lambda$CDM in a lower redshift range where SN are numerous. Nevertheless, as higher redshift QSOs are added, QSOs gradually return larger values of $\Omega_{m}$ until one enters an Einstein-de Sitter Universe (EdS) (spatially flat FLRW with only pressureless matter) when $z \gtrsim 1$. Taken at face value, QSOs transition from a dark energy (DE) dominated Universe to a matter dominated Universe, which may partially explain the preference of QSO data for larger values of $\Omega_m$, and consequently less DE \cite{Yang:2019vgk, Velten:2019vwo}. 

Next, we show hints of the same evolution of $\Omega_{m}$ with redshift, but in Type Ia SN \cite{Pan-STARRS1:2017jku}. Concretely, we show that as $\Omega_{m}$ increases, then $H_0$ decreases, at least within the flat $\Lambda$CDM model. This trend is simply recovering earlier results in the literature \cite{Dainotti:2021pqg, Horstmann:2021jjg, Dainotti:2022bzg}. The main point is that we see the \textit{same} trend independently in both QSOs and SN, both of which have distinct strengths and weaknesses. On one hand, QSOs are plentiful at higher redshifts and have good statistics, but are relatively new cosmological probes (see Ref. \cite{Moresco:2022phi} for a review) and suffer from greater intrinsic scatter. On the flip side, SN represent a cornerstone of modern cosmology, but become sparse at higher redshifts, thereby preventing us from confirming that $\Omega_{m} > 0.3$. Nevertheless, combining both probes, not only does one recover a Planck-$\Lambda$CDM Universe in a similar redshift range $z \lesssim 0.7$, but one sees hints of a deviation from the Planck-$\Lambda$CDM at $z \sim 1$. 

In a bid to assign a statistical significance to SN observations, we note that fits of higher redshift mock $\Lambda$CDM data lead to distributions with non-Gaussian tails towards larger $\Omega_m$ and smaller $H_0$ values. We explain this feature as an inherent bias in flat $\Lambda$CDM, which makes it more likely that early Universe determinations of $H_0$ and $S_8$ are smaller and larger, respectively, {than late Universe counterparts}. Interestingly, strong lensing time delay also reports a  descending trend in $H_0$ with lens redshift \cite{Wong:2019kwg, Millon:2019slk}, prompting Ref. \cite{Krishnan:2020obg} to investigate the same trend in other cosmological probes. We also encounter some intriguing trends in $\Omega_m$ with BAO observations, which we present in supplemental material. In some sense, $\Lambda$CDM may be a smart model that predicts its own demise, including a decreasing $H_0$ with redshift (see Refs. \cite{Krishnan:2020vaf, Krishnan:2022fzz} for related comments). Given the universal confidence in SN cosmology, the outlined trends can be confirmed or refuted by simply increasing the number of high redshift ($z \sim 1$) SN. This is expected to happen soon \cite{Scolnic:2019apa}. 

\section{QSOs}
Standardizable QSOs represent a game changer for cosmology. They are plentiful, in contrast to gamma-ray bursts (GRBs) \cite{Amati:2008hq, Cardone:2009mr, Cardone:2010rr, Amati:2013sca, Dainotti:2013fra, Postnikov:2014aua, Cucchiara:2011pj}, but like GRBs, promise to open up the redshift range beyond SN.
%A number of standardization methods have been proposed for quasars such as the variability method discussed in \cite{Solomon:2021jml} and the Risaliti and Lusso method relation \cite{Risaliti:2015zla, Risaliti:2018reu}.
%The work of \cite{Solomon:2021jml} is still young in its application so we focus on the Risaliti and Lusso method.
Based on an empirical relation between X-ray and UV QSO fluxes, Risaliti and Lusso have proposed a relation intrinsic to QSO luminosities for cosmological purposes (\ref{RL}) (see \cite{Watson:2011um, Wang:2013ha, LaFranca:2014eba,Solomon:2021jml} for other methods). The constant $\gamma$ is directly inherited from the flux relation through the standard luminosity-flux relation, $L = 4 \pi D_{L}(z)^2 F$, where $D_{L}(z)$ denotes the luminosity distance. The robustness of $\gamma \approx 0.6$ to redshift evolution has been demonstrated over both orders of magnitude in luminosity and extended redshifts \cite{Vignali:2002ct, Just:2007se, Lusso:2009nq, Salvestrini:2019thn, Bisogni:2021hue}.  

However, in contrast to SN, there is considerable intrinsic scatter in QSO fluxes. As with SN in the 1990s \cite{Phillips:1993ng}, before corrections for color, shape and host galaxy mass \cite{Tripp:1997wt, NearbySupernovafactory:2013qtg} (however, see \cite{NearbySupernovaFactory:2018qkd, Kang:2019azh, Jones:2018vbn, Lee:2021txi} for ongoing debate), this scatter necessitates an additional intrinsic dispersion parameter $\delta$. Given corrections made to SN since the 1990s, it is worth bearing in mind that (\ref{RL}) is a working proposal and future corrections may be necessary, especially in light of criticisms \cite{Khadka:2020whe, Khadka:2020vlh, Khadka:2021xcc}. Moreover, one cannot rule out the possibility going forward that better data selection criteria could also reduce the scatter. Nevertheless, we adopt the Risaliti-Lusso relation (\ref{RL}) and obtain best fit parameters by marginalizing or maximizing the likelihood function \cite{Risaliti:2015zla, Risaliti:2018reu}, 
\be
\label{L1}
\mathcal{L} = - \frac{1}{2} \sum_{i=1}^{N} \left[ \frac{ \left(\log_{10} F^{\textrm{obs}}_{X, i} - \log_{10} F^{\textrm{model}}_{X,i}\right)^2}{s^2_{i}} + \ln (2 \pi s_i^2) \right], 
\ee 
where the cosmological model enters through the flux relation that follows from (\ref{RL}):
\be
\label{fluxes}
\log_{10} F_{X} = \beta  + \gamma \log_{10} F_{UV} +  (\gamma-1)  \log_{10} (4\pi D_{L}^2).  
\ee
Here, $s_i^2 = \sigma_i^2+ \delta^2$ in (\ref{L1}) contains the measurement error on the observed flux $\log_{10} F^{\textrm{obs}}_{X,i}$. The $F_{UV}$ errors are ignored \cite{Risaliti:2015zla}. Note that $F_{X}$ and $F_{UV}$ errors are considerably smaller than $\delta$.

\begin{figure}[htb]
   \centering
\includegraphics[width=85mm]{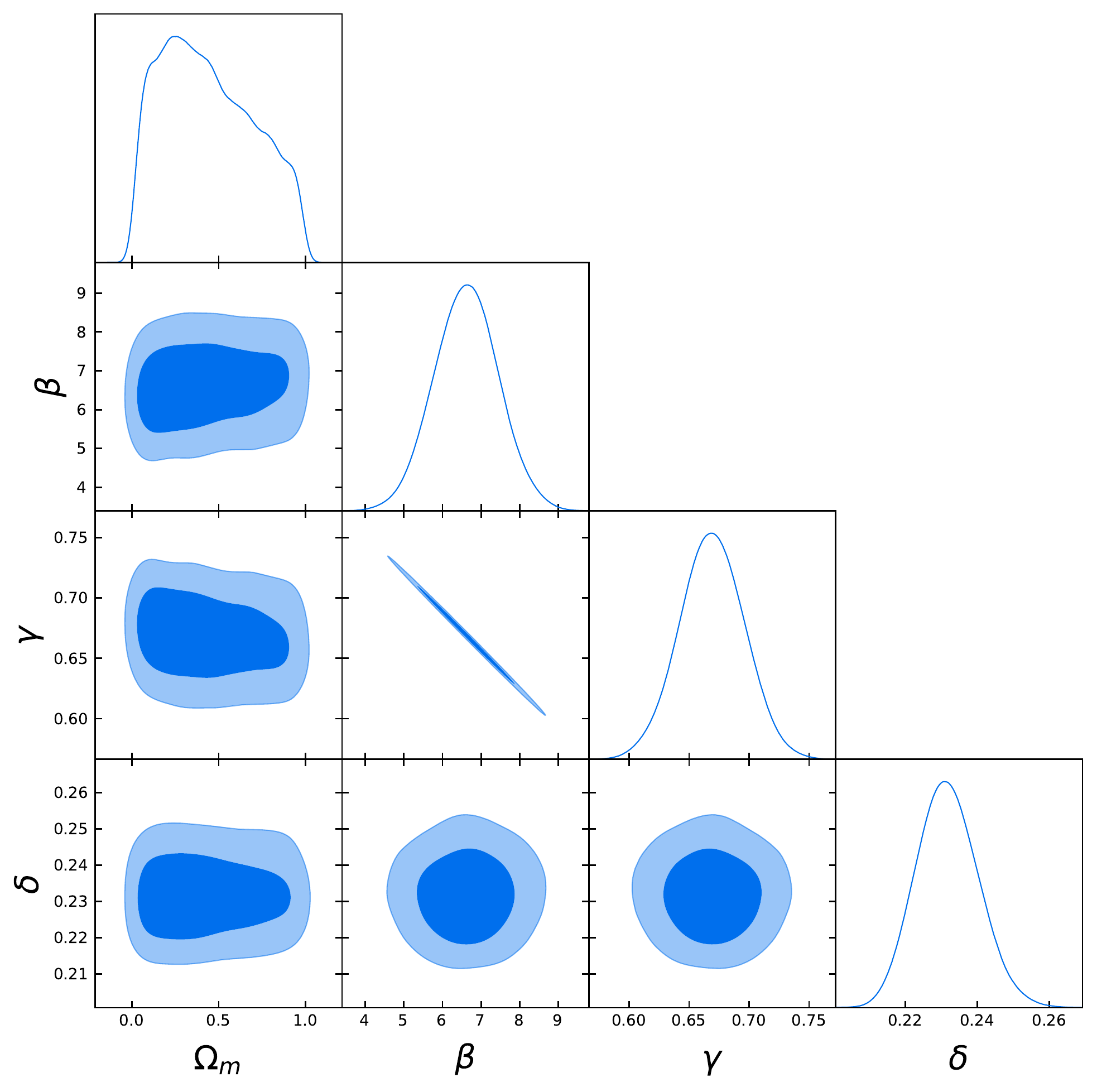}
\caption{Marginalized parameters for the QSO sample \cite{Lusso:2020pdb} with cutoff $z_{\textrm{max}} = 0.7$.}
\label{zmax0.7} 
\end{figure}

While it is customary in the literature to calibrate QSOs with SN to identify $\beta$ \cite{Risaliti:2015zla, Risaliti:2018reu}, this risks hiding physics that is intrinsic to QSOs, since QSOs simply track SN, so here we work with uncalibrated QSOs and flat $\Lambda$CDM with nuisance parameters $(\beta, \gamma, \delta)$. Since $\beta$ is degenerate with $H_0$, one cannot determine both, so we fix $H_0 = 70$ km/s/Mpc. Our first goal is to restrict the maximum redshift $z_{\textrm{max}}$ of the latest sample of 2421 QSOs \cite{Lusso:2020pdb} in order to demonstrate that QSOs at lower redshifts, where SN are numerous, inhabit a Planck-$\Lambda$CDM Universe with $\Omega_{m} \approx 0.3$. In Fig. \ref{zmax0.7}, we confirm that matter density is peaked close to the Planck value $\Omega_{m} \approx 0.3$ when $z_{\textrm{max}} = 0.7$ and there are 398 QSOs in the range. Therefore, in the redshifts where they overlap well, both SN and QSOs agree on DE, in contrast to findings \cite{Yang:2019vgk, Velten:2019vwo} over extended redshift ranges. Note, in contrast to Refs. \cite{Risaliti:2015zla, Risaliti:2018reu}, here the QSOs are uncalibrated, so they recover DE without guidance from SN. This is easy to take for granted, but it is a valid consistency test for the QSOs.  

\begin{table}[htb]
\centering 
\begin{tabular}{c|c|c|c}
 \rule{0pt}{3ex} $z_{\textrm{max}}$ & $ \Omega_{m}$ & $ \beta$ & $\gamma$ \\
\hline 
\rule{0pt}{3ex} \multirow{2}{*}{$0.7$ (398 QSOs)} & $0.266$ & $6.601$  & $0.670$ \\
\rule{0pt}{3ex}  & $0.411^{+0.342}_{-0.259}$ & $6.620^{+0.814}_{-0.841}$ & $0.669^{+0.027}_{-0.027}$  \\
\hline
\rule{0pt}{3ex} \multirow{2}{*}{$0.8$ (543 QSOs)} & $0.418$ & $7.162$  & $0.652$ \\
\rule{0pt}{3ex}  & $0.511^{+0.305}_{-0.275}$ & $7.162^{+0.715}_{-0.712}$ & $0.651^{+0.023}_{-0.023}$  \\
\hline
\rule{0pt}{3ex} \multirow{2}{*}{$0.9$ (678 QSOs)} & $0.592$ & $7.736$  & $0.633$ \\
\rule{0pt}{3ex}  & $0.601^{+0.248}_{-0.250}$ & $7.709^{+0.662}_{-0.679}$ & $0.633^{+0.022}_{-0.021}$  \\
\hline
\rule{0pt}{3ex} \multirow{2}{*}{$1$ (826 QSOs)} & $0.953$ & $7.921$  & $0.626$ \\
\rule{0pt}{3ex}  & $0.717^{+0.184}_{-0.231}$ & $7.792^{+0.571}_{-0.571}$ & $0.631^{+0.019}_{-0.019}$  \\
\end{tabular}
\caption{Best fit and marginalized inferences of ($\Omega_{m}, \beta, \gamma)$ for QSOs below a maximum redshift $z_{\textrm{max}}$. }
\label{QSOvsZ}
\end{table}

Now comes a remarkable observation.
Namely, as the maximum redshift ticks up towards $z=1$, the best fit and marginalized values of $\Omega_{m}$ also increase towards $\Omega_{m} \approx 1$ in the flat $\Lambda$CDM model.
This can be seen from Table \ref{QSOvsZ}, where we have omitted $\delta$ as it is consistent with $\delta \sim 0.23$ throughout. Since we have imposed the flat prior $ 0 < \Omega_{m} < 1$, our marginalized results are impacted by the bounds, but we have checked that the best fit values for $\Omega_m$ agree with the peaks of the $\Omega_{m}$ distribution.
It should be noted that we have made use of few inputs, merely that (\ref{RL}) holds and we marginalize or maximize the likelihood (\ref{L1}) following the Risaliti-Lusso prescription \cite{Risaliti:2015zla, Risaliti:2018reu}.
Nevertheless, we recover a Planck-$\Lambda$CDM Universe, where it is expected, in more or less the same redshift range as SN, however QSOs transition to an EdS Universe ($\Omega_m=1$) with larger $z_{\textrm{max}}$. Concretely, at $z_{\textrm{max}} \approx 1.3$, the QSOs inhabit an EdS Universe. Throughout, we find that $\gamma \gtrsim 0.6$ even as $z_{\textrm{max}}$ is increased beyond $z_{\textrm{max}} \approx 1.3$.

Note, as explained in \cite{Lusso:2020pdb}, there is concern that some of the UV fluxes have been extrapolated from the optical below $z = 0.7$, however, as we have seen, QSOs still recover DE. Moreover, as is evident from Table \ref{QSOvsZ}, there is evolution in ($\beta, \gamma$) as the redshift range is extended. One could seize upon this fact and immediately jump to the conclusion that QSOs are not standardizable, but there is a kicker; SN show the same evolution in the central value of $\Omega_{m}$. Moreover, as we will argue later, evolution in $\Omega_{m}$ with redshift may be fundamental to the flat $\Lambda$CDM Hubble diagram, and the remaining parameters simply compensate. Thus, if $\Omega_m$ evolves, so too must $\beta$ or $\gamma$ (cf. comments in \cite{Khadka:2020whe, Khadka:2020vlh, Khadka:2021xcc}). We will see the same with SN, where $H_0$ compensates evolution in $\Omega_m$. In supplemental material, we discuss the robustness of the QSO results to subsample restrictions.  

Finally, our analysis here can be contrasted with the methodology in Ref. \cite{Dainotti:2022rfz}, where a fiducial cosmology and corresponding Hubble diagram are assumed, while the luminosities are corrected for redshift evolution. Here we are conversely interested in extracting the cosmology, in particular $\Omega_m$, so the results in Table \ref{QSOvsZ} assume the Risaliti-Lusso relation (\ref{RL}). For this reason, some differences in the values of $(\beta, \gamma)$ are expected, especially here since $(\beta, \gamma)$ compensate for evolution in $\Omega_m$, as explained above.

\section{Pantheon SN}
\begin{figure}[htb]
   \centering
\begin{tabular}{c}
\includegraphics[width=90mm]{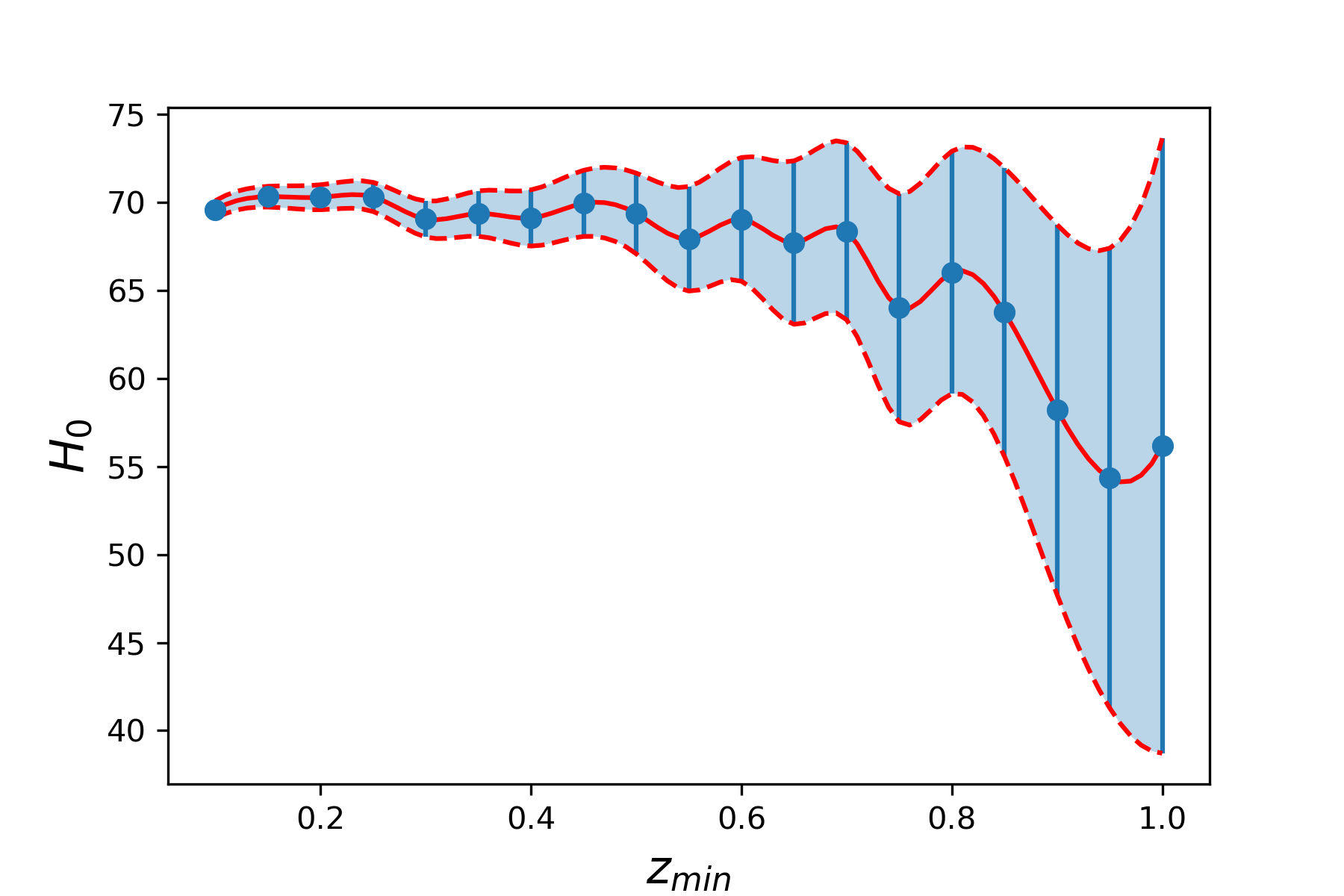} \\
\includegraphics[width=90mm]{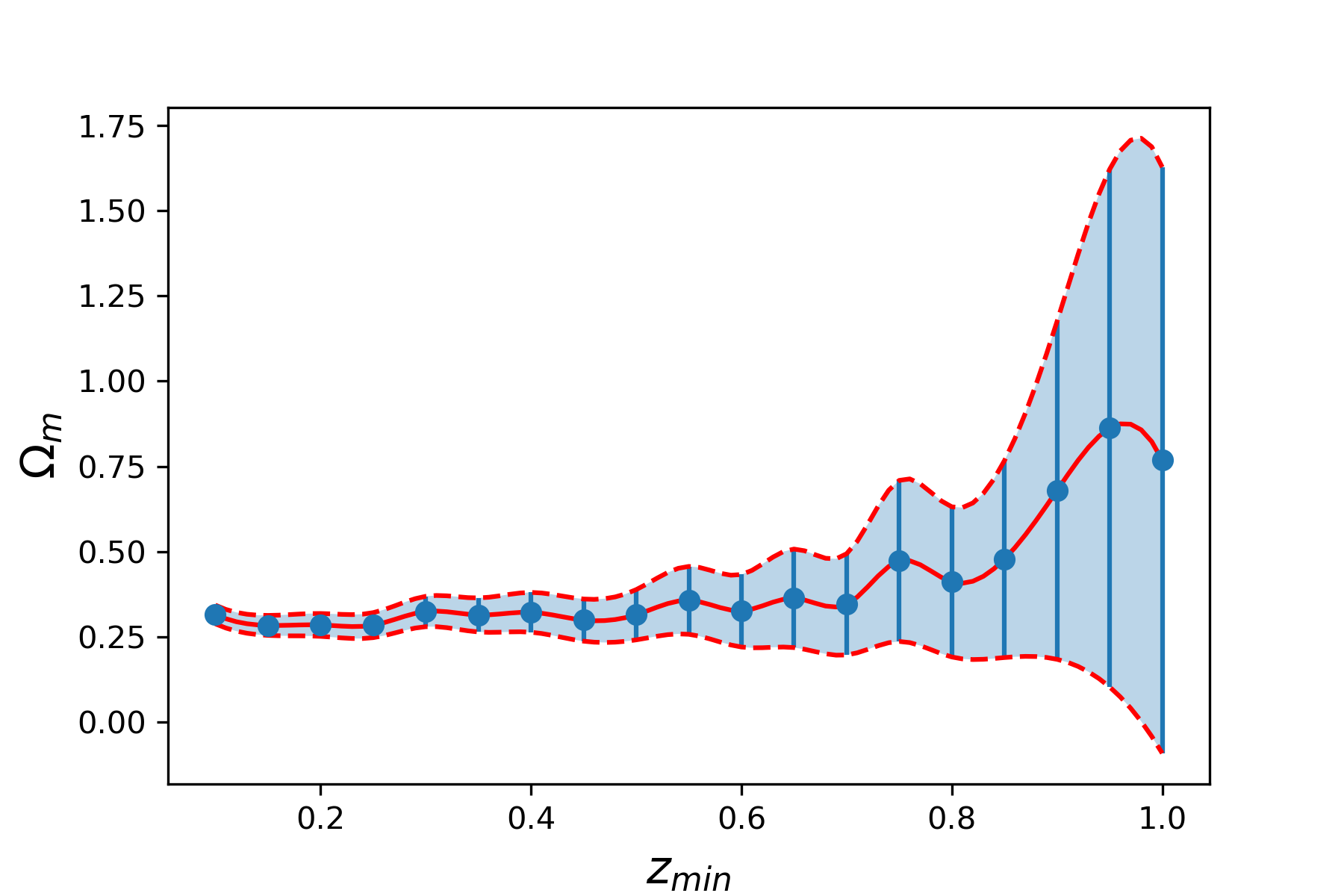}
\end{tabular}
\caption{Variations of best fit cosmological parameters $(H_0, \Omega_{m})$ as low redshift SN below $z_{\textrm{min}}$ are removed. The central values of $H_0$ and $\Omega_{m}$ favour lower and higher values at higher redshifts, respectively. The error bars denote the variance in fitted $H_0$ and $\Omega_m$ values at each $z_{\textrm{min}}$ taken before performing a cubic interpolation.}
\label{H0_om_z} 
\end{figure}

We now switch gears to Pantheon SN \cite{Pan-STARRS1:2017jku}, where it is already documented that $H_0$ descends \cite{Dainotti:2021pqg, Dainotti:2022bzg} and $\Omega_{m}$ increases with redshift binning \cite{Horstmann:2021jjg} (see their Fig. 6). Here we simply confirm these results by imposing a low redshift cutoff $z_{\textrm{min}}$, which allows us to decouple SN below a given redshift.  
% Concretely, we the absolute magnitude fix $M_{B} = -19.35$ mag,
For concreteness, we fix the absolute magnitude to $M_{B} = -19.35$, which is consistent with a nominal $H_0 \approx 70$ km/s/Mpc value, while fitting $H_0$ and $\Omega_{m}$ within the flat $\Lambda$CDM model in intervals of $\Delta z = 0.05$ in the redshift range $ 0.1 \leq z \leq 1$. We show the results of this exercise in Fig. \ref{H0_om_z}, where we include $1 \sigma$ confidence intervals and interpolate between the values of cosmological parameters using a cubic spline. Note, our analysis includes both statistical and systematic uncertainties through the full Pantheon covariance matrix, which we crop appropriately when we remove SN.

\begin{figure}[htb]
   \centering
\includegraphics[width=90mm]{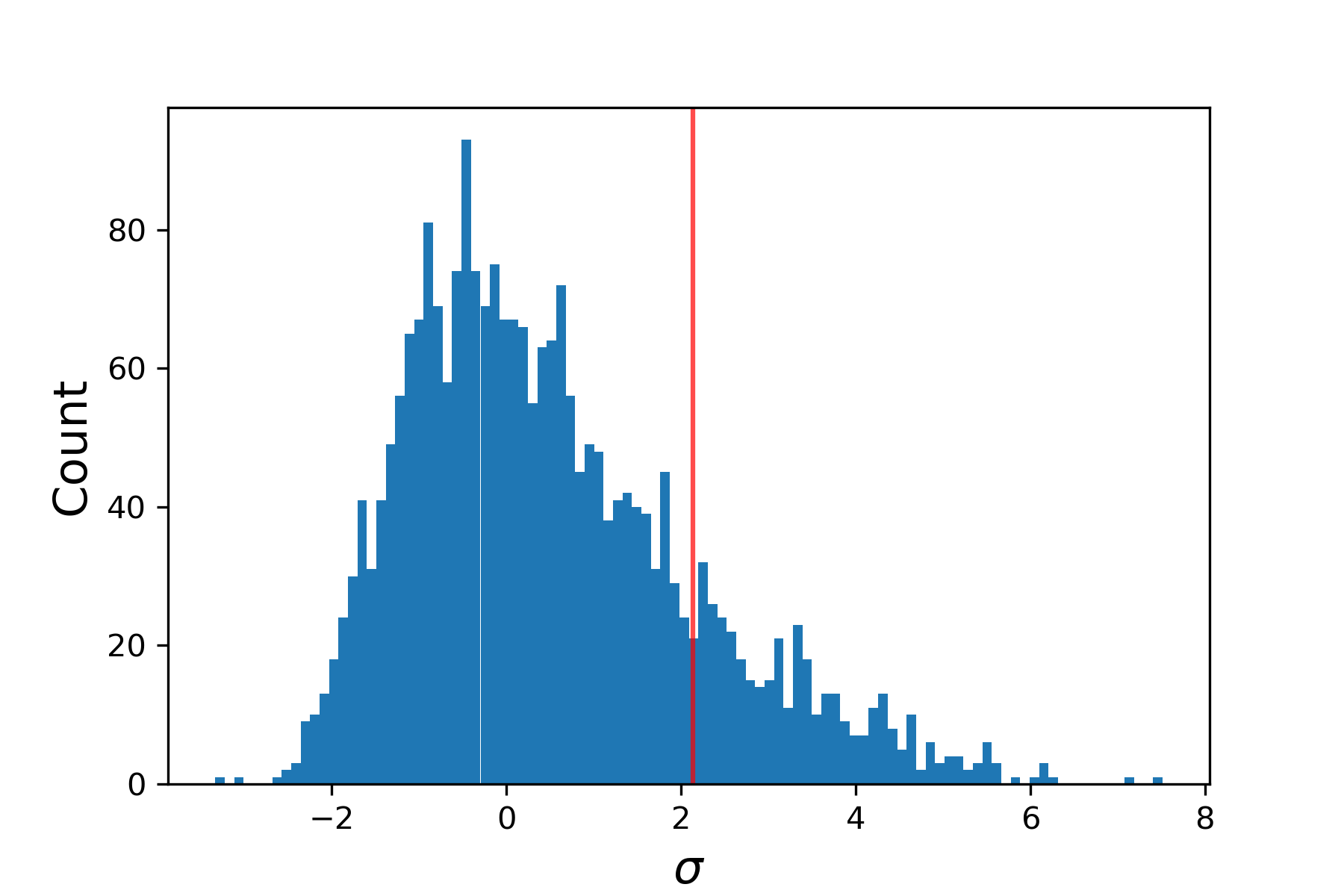}
\caption{$\sim 2500$ mock realisations of the Pantheon SN sample and the corresponding sum (\ref{sigma}). The value corresponding to real data is denoted by the red line.}
\label{sum} 
\end{figure}

While the result is expected \cite{Dainotti:2021pqg, Dainotti:2022bzg, Horstmann:2021jjg}, it is interesting to note that Pantheon+ shows a similar $\Omega_m$ trend with $z_{\textrm{min}}$ through to $z_{\textrm{min}}= 0.3$ \cite{Brout:2022vxf} (see their Fig. 16). The descending $H_0$ trend is also reminiscent of similar trends in strong lensing time delay with statistical significance $1.9 \sigma$ \cite{Wong:2019kwg} and $1.7 \sigma$ \cite{Millon:2019slk}, respectively.\footnote{In strong lensing time delay, one is less sensitive to $\Omega_m$.} We performed approximately 2500 simulations of mock data based on Planck values, where we kept track of the sum of the discrepancy with Planck \cite{Planck:2018vyg} in $\Omega_{m}$ evaluated at each $z_{\textrm{min}}$ we sampled, 
\be
\label{sigma}
\sigma:= \sum_{z_{\textrm{min}}} (\Omega^{z_{\textrm{min}}}_{m} - \Omega_{m}^{\textrm{Planck}}). 
\ee
One could define an analogous sum for $H_0$, but since $(H_0, \Omega_m)$ are anti-correlated, it suffices to focus on one parameter. For the real data, this sum is positive, $\sigma = 2.14$, as is evident from Fig. \ref{H0_om_z}. We present the simulations in Fig. \ref{sum}, where we find that larger positive sums arise by chance with probability $p = 0.16$ ($\sim 1\sigma$), which is consistent with the $1 \sigma$ deviation from Planck-$\Lambda$CDM evident with real data in Fig. \ref{H0_om_z}. 

Interestingly, we find that the median and $1 \sigma$ confidence intervals are all shifted to larger $\sigma$ values. In particular, we find that the median is $\sigma = 0.19$, while the $1 \sigma$ confidence interval is $-1.06 < \sigma < 2.16$. We will argue in the next section that this is an intrinsic feature of the flat $\Lambda$CDM model, which arises at higher redshifts. However, here it is not clear how much of this effect is attributable to observations and how much to the $\Lambda$CDM model. Either way, there is a problem. That point aside, the goal here is simply to point out that SN are expected to follow QSOs, if QSOs are \textit{bona fide} standardizable candles.

%\begin{table}[htb]
%\centering 
%\begin{tabular}{c|c|c|c}
%Sample & SN Count & $H_0$ (km/s/Mpc)& $\Omega_m$ \\
%\hline 
%Low z & $172$ & $70.23^{+0.81}_{-0.72}$ & %$0.613^{+0.266}_{-0.326}$ \\
%SDSS & $335$ & $69.16^{+0.73}_{-0.74}$ & %$0.331^{+0.070}_{-0.066}$ \\
%PS1 & $279$ & $68.48^{+0.76}_{-0.77}$ & %$0.382^{+0.057}_{-0.054}$ \\
%SNLS & $236$ & $69.94^{+1.04}_{-1.13}$ & %$0.279^{+0.043}_{-0.041}$ \\
%HST & $26$ & $70.78^{+7.18}_{-8.10}$ & %$0.314^{+0.243}_{-0.136}$ \\
%\end{tabular} 
%\caption{\red{This table will only confuse people, so I will take it out. Any objections?}}
%\label{table1}
%\end{table} 
\section{$\Lambda$CDM Digression} 
As is evident from Fig. \ref{sum}, the sum distribution is not Gaussian and has developed some non-Gaussian tails. Here we will argue that these tails are a generic feature of the flat $\Lambda$CDM model that arise at higher redshifts. See \cite{Colgain:2022rxy} where these ideas are further developed. To begin, recall the flat $\Lambda$CDM model: 
\be
\label{LCDM}
H(z)^2 = H_0^2 (1 - \Omega_{m}) + H_0^2\Omega_{m} (1+z)^3. 
\ee
Here, the Hubble constant $H_0$ is an integration constant from the perspective of the Friedmann equations, while $\Omega_{m}$ is the matter density today. The latter is bounded in a physical regime, $0 < \Omega_{m} \leq 1$; $\Omega_m=0$ is ruled out by the mere fact that $H(z)$ is not a constant and $\Omega_m=1$ corresponds to the EdS Universe.    

At low redshifts, $z \ll 1$, expanding (\ref{LCDM}), one has $H(z) = H_0 \left( 1 + \frac{3}{2} \Omega_m z + O(z^2) \right)$. Thus low redshift data first constrains $H_0$ and then $\Omega_{m}$, which is subleading in $z < 1$. Within the prevailing Planck-$\Lambda$CDM Universe \cite{Planck:2018vyg}, one expects $\Omega_{m} \approx 0.3$. However, as is clear from (\ref{LCDM}), the high redshift behaviour of the Hubble parameter is $H(z) \sim H_0 \sqrt{\Omega_{m}} (1+z)^{\frac{3}{2}}$, which only depends on a single parameter $H_0^2\Omega_m$. Thus, high redshift observational data ensures that $H_0$ and $\Omega_m$ are anti-correlated; as $H_0$ increases, $\Omega_m$ decreases, and \textit{vice versa}. Observe that neglecting galaxy BAO, the anti-correlation between $H_0$ and $\Omega_m$ is pretty generic \cite{Lin:2019htv} (see their Fig. 1). Note, we have dropped the $(1- \Omega_{m})$ term as despite being relevant at lower redshifts, it becomes less relevant at higher redshifts. As explained in \cite{Colgain:2022rxy}, there is an inevitable spreading in the $H_0^2 (1-\Omega_m)$ distribution of best fit values within the flat $\Lambda$CDM model in high redshift bins, which pushes best fit $\Omega_m$ values away from the Planck value $\Omega_m \sim 0.3$ and towards the boundary 
%ies $\Omega_m = 0$ and 
$\Omega_m = 1$. This is a direct consequence of the irrelevance of DE density at higher redshifts. 
%Now, since data treats less relevant terms by setting them to zero, of course within errors, we can infer that high redshift data sets $(1- \Omega_m)$ to zero in (\ref{LCDM}) by gradually increasing $\Omega_{m} \rightarrow 1$ as the effective redshift $z_{\textrm{eff}}$ of data increases. Finally, since the combination $H_0^2 \Omega_{m}$ is fixed, this leads to a decreasing trend in $H_0$ with $z_{\textrm{eff}}$.  

\begin{figure}[htb]
   \centering
\begin{tabular}{c}
\includegraphics[width=90mm]{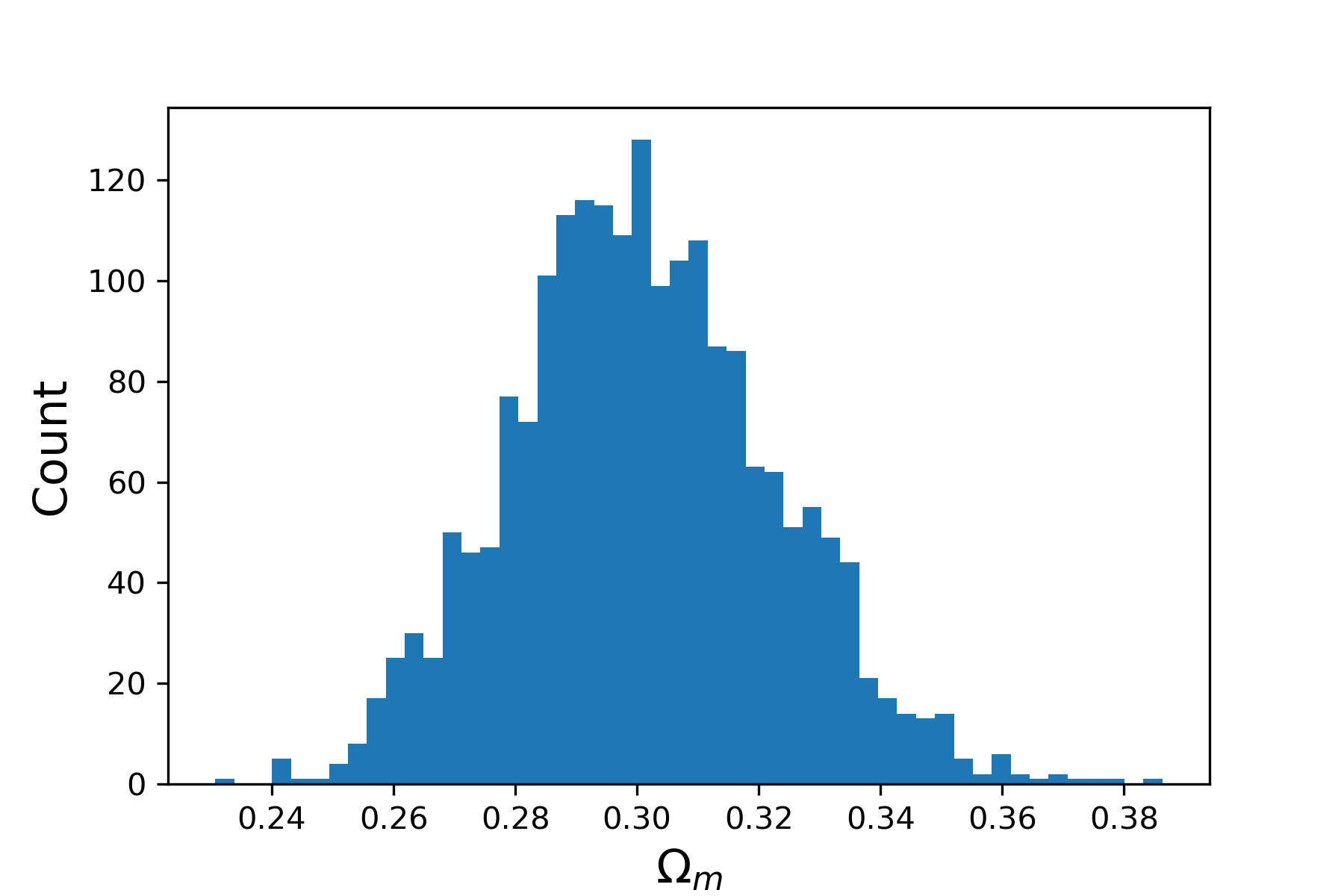} \\
\includegraphics[width=90mm]{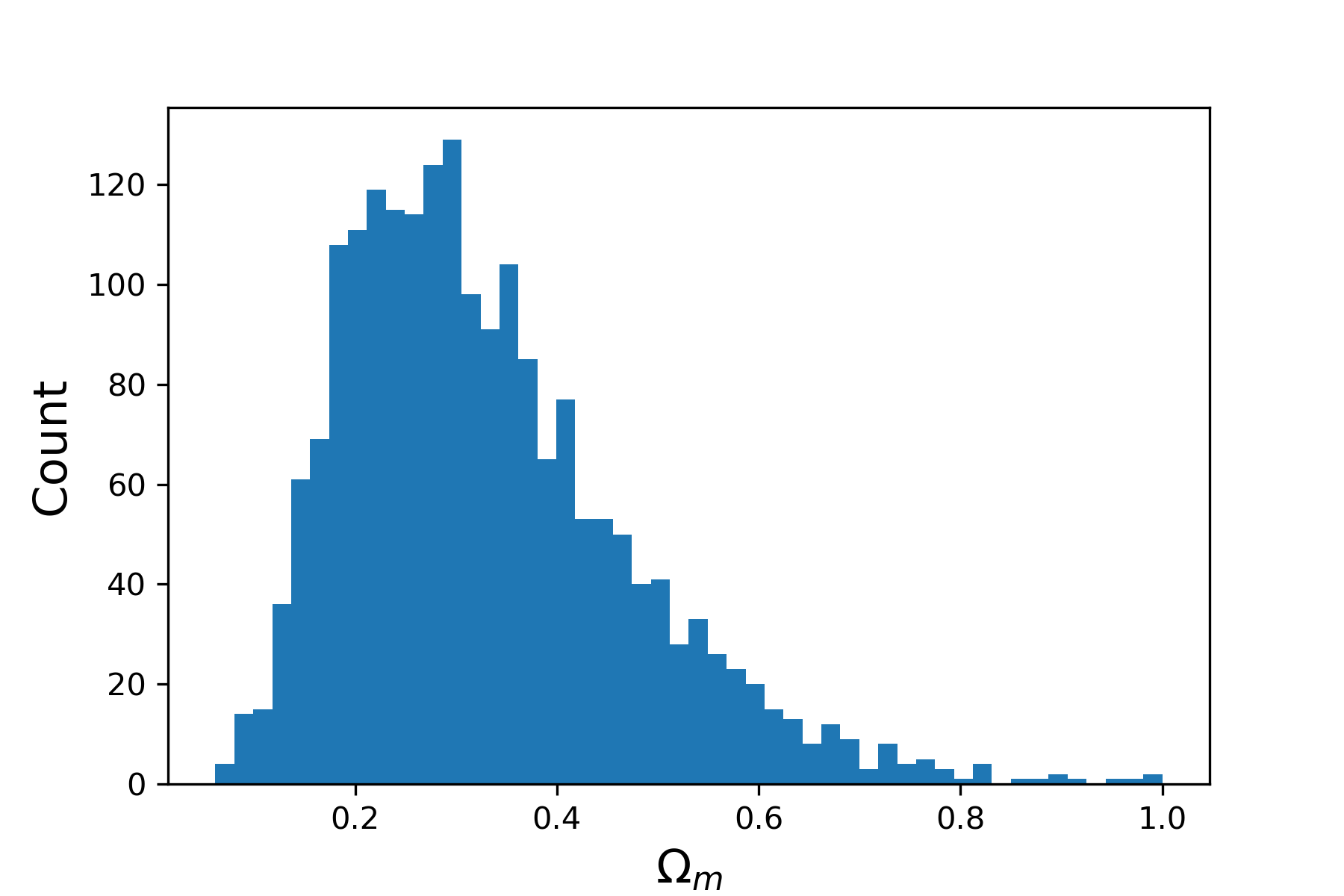}
\end{tabular}
\caption{The distribution of best fit $\Omega_m$ for 2000 mocks of the full Pantheon SN sample with redshifts $0 < z \leq 2.26$ (above) and SN subsample with redshifts $ 0.7 < z \leq 2.26$ (below). The $\Omega_{m}$ distribution becomes non-Gaussian at higher redshifts.}
\label{omdist} 
\end{figure}

The pertinent question now is, how strong is this bias and when does it become a concern? In particular, could it explain the effect that we see in Fig. \ref{H0_om_z}? Once again, we turn to SN mocks, but now instead of summing, we simply work with the full sample of 1048 SN and and a subsample of 124 SN above $z = 0.7$. The effective redshifts are $z_{\textrm{eff}} \approx 0.28$ and $z_{\textrm{eff}} \approx 0.9$, respectively, where we have weighted by the uncertainty in apparent magnitude $m_{B}$. Here we have chosen values that lead to an exaggerated effect, but for values of $z_{\textrm{eff}}$ in between, one still notices some effect. For both the full sample and subsample, we mock up SN data with canonical values $H_0 = 70$ km/s/Mpc and $\Omega_{m} = 0.3$. In total, we produce 2000 mock realisations of the data and fit the flat $\Lambda$CDM model back to each mock and record the best fit values of the cosmological parameters. As can be seen from Fig. \ref{omdist}, the distribution of best fit values of $\Omega_{m}$ develops a long tail for larger $\Omega_{m}$ values at higher redshifts. Although we omit the plot, it is a given that the $H_0$ distribution shows a similar tail towards smaller values of $H_0$ (see \cite{Colgain:2022rxy}). That being said, we have checked that both the mean and median are consistent with the input values for $H_0$ and $\Omega_m$, which simply underscores that one is analyzing mock data. The real story here is the high redshift tails.        

It is an easy deduction to see that the non-Gaussian distribution in our sum in Fig. \ref{sum} is coming from the higher redshift contributions to the sum. {As teased out in \cite{Colgain:2022rxy}, the non-Gaussian tails at higher redshifts arise from the spreading of $H_0^2 (1-\Omega_m)$ distribution of best fits until one encounters the boundary at $\Omega_m =1$. This boundary precludes negative DE densities in the flat $\Lambda$CDM model. Thus, as one bins data by redshift and confronts with the flat $\Lambda$CDM model, non-Gaussian tails in the direction of larger $\Omega_m$ and lower $H_0$ values arise. This feature, which is evident in mocks, and therefore inherent to the flat $\Lambda$CDM model, suggests that observations of decreasing $H_0$ values \cite{Wong:2019kwg, Millon:2019slk, Krishnan:2020obg, Dainotti:2021pqg, Dainotti:2022bzg, Horstmann:2021jjg} with redshift are physical and can be expected within flat $\Lambda$CDM.}
%We have already provided the intuitive argument why this should be the case. 
%In the flat $\Lambda$CDM model, as data quality reduces through binning etc. at higher redshifts, the DE term, which is proportional to $1-\Omega_m$ becomes less relevant. As a result, this term can be dropped by increases in $\Omega_m$ and decreases in $H_0$. \textit{This is an inherent feature of the flat $\Lambda$CDM model} and explains why observations of decreasing $H_0$ values \cite{Wong:2019kwg, Millon:2019slk, Krishnan:2020obg, Dainotti:2021pqg, Dainotti:2022bzg, Horstmann:2021jjg} with redshift may be expected. 
Note, our analysis here has been model dependent within $\Lambda$CDM, but there are diagnostics allowing one to track trends model independently \cite{Krishnan:2020vaf, Sahni:2008xx}.

\section{Discussion}
Although we have glossed over a host of interesting details, let us revisit the facts. Risaliti \& Lusso have a proposal for standardizable QSOs \cite{Risaliti:2015zla, Risaliti:2018reu}, based on the relation (\ref{RL}), which one can argue is intrinsic to QSOs \cite{Dainotti:2022rfz}. In turn, QSOs recover the Planck-$\Lambda$CDM Universe at lower redshits $z \lesssim 0.7$, in line with the expectations of SN cosmology. Nevertheless, SN and QSOs are very different beasts and while SN are weighted towards low reshifts, $z_{\textrm{eff}} \approx 0$, the Risaliti-Lusso QSO samples are more numerous at higher  redshift, $z_{\textrm{eff}}\approx 1$. We have demonstrated that within the Risaliti-Lusso assumptions QSOs transition from a Planck-$\Lambda$CDM Universe to an EdS Universe as one increases the redshift range. 

One could write off this behaviour simply on the grounds that QSOs are not standardizable, but what then if SN show similar trends? As we have shown, there is an increasing $\Omega_{m}$, decreasing $H_0$ trend, in Pantheon SN \cite{Scolnic:2019apa} as the low redshift SN anchoring the sample in the DE dominated regime are decoupled (see also \cite{Dainotti:2021pqg, Dainotti:2022bzg, Horstmann:2021jjg}). Note, while the QSOs become more numerous at higher redshifts, the SN become less numerous, and statistics currently prevent a definitive statement. This will change in coming years and the Roman Space Telescope \cite{Spergel:2015sza, Hounsell:2017ejq} is expected to lead to $\times 1000$ improvement in $z > 1$ SN statistics. This will allow us to confirm if both QSOs and SN are following the same trend. It is worth stressing that any evolution in $H_0$ within SN is equivalent to evolution in absolute magnitude $M_B$, so if confirmed, it represents a stark choice between SN cosmology and the flat $\Lambda$CDM model. 

Finally, we have explained why this trend  to be expected in the flat $\Lambda$CDM model. The ideas are further developed in  \cite{Colgain:2022rxy}. In short, there is no guarantee that $\Omega_m$ is not increasing and $H_0$ is not  decreasing at higher redshifts as one bins the data. Indeed, it is possible that our real SN sample in Fig. \ref{H0_om_z} is somewhere in the tails of Fig. \ref{omdist}. The non-Gaussian tail highlights the ease at which one could perform an experiment and get higher values of $\Omega_m$ and lower values of $H_0$. Ultimately, this suggests that documented trends in $H_0$ \cite{Wong:2019kwg, Millon:2019slk, Krishnan:2020obg, Dainotti:2021pqg, Dainotti:2022bzg, Horstmann:2021jjg} in the literature may be physical. Moreover, as we discuss in supplemental material, an increasing $\Omega_m$ with $z_{\textrm{eff}}$ may be supported by BAO observations \cite{BOSS:2016wmc, Bautista:2020ahg, Gil-Marin:2020bct}, but this requires further investigation.  

Observe that this also gives a new perspective on cosmological or $\Lambda$CDM tensions. All things being equal, one is more likely to find that $H_0$ is lower at higher redshifts, thereby seemingly explaining why early Universe determinations of $H_0$ (and $\Omega_m$) are indeed smaller (and larger) when one interprets the physical Universe through the $\Lambda$CDM model (see  \cite{Wagner:2022etu} for related comments).  Moreover, as $\Omega_m$ increases, so too does $S_8$ within the flat $\Lambda$CDM model (see Fig. 1 of \cite{Nunes:2021ipq}). Once again, this trend could explain why Planck measures larger values of $S_8$. 

Going further, there is a lensing anomaly in the CMB and it is well documented that one infers a lower $H_0$ and higher $\Omega_m$ from higher multipoles \cite{Addison:2015wyg}. Could this too be explained as some artifact of viewing CMB through the prism of flat $\Lambda$CDM? In addition, could any preference in data sets for interacting DE models \cite{Gavela:2009cy, DiValentino:2017iww} be explained by this trend? Regardless, there is an inherent bias in the flat $\Lambda$CDM model, as the non-Gaussian tails in SN mocks demonstrate, and whether larger $\Omega_m$ values come from this bias or the physical data is less relevant. Evidently, SN (and perhaps BAO) have the potential to shore up Risaliti-Lusso QSOs \cite{Risaliti:2015zla, Risaliti:2018reu} as standardizable candles while ruling out the Planck-$\Lambda$CDM Universe. On the flip side, if $\Omega_m$ does not increase with redshift in SN and BAO, then the intrinsic scatter in QSOs is presumably problematic. Attention must then focus on reducing the scatter or turning to other approaches for standardizable QSOs \cite{Watson:2011um, Wang:2013ha, LaFranca:2014eba,Solomon:2021jml}. 

\section{Acknowledgements} 
We thank Stephen Appleby, Eleonora Di Valentino, Dragan Huterer, Chethan Krishnan, Ziad Sakr, Jenny Wagner and Kenneth Wong for discussion and comments on earlier drafts. We also credit an anonymous referee at PRD for inspiring this letter. MGD acknowledges the Division of Science and NAOJ for the support. GB and SC acknowledge the Istituto Nazionale di Fisica Nucleare (INFN), sezione di Napoli, \textit{iniziative specifiche} QGSKY and MOONLIGHT-2. E\'OC was supported by the National Research Foundation of Korea grant funded by the Korea government (MSIT) (NRF-2020R1A2C1102899). DS is partially supported by the US National Science Foundation, under Grant No. PHY-2014021. MMShJ would like to acknowledge SarAmadan grant No. ISEF/M/400121.

\appendix 

\section{Further QSO Checks}
Here we perform a further check on the increasing trend of $\Omega_{m}$ with redshift highlighted in the text. In particular, it has been argued in Ref. \cite{Khadka:2021xcc} that the SDSS-4XMM subsample (1644 QSOs from 2421) of the QSO sample \cite{Lusso:2020pdb} is the most untrustworthy, since it leads to a Risaliti-Lusso relation that varies with both cosmological models and redshift. In Table \ref{QSOvsZ_subsample} we repeat the exercise without the SDSS-4XMM subsample to show that the increasing trend of $\Omega_{m}$ with $z_{\textrm{max}}$ persists with the removal of the SDSS-4XMM subsample. Note, the cosmological parameters differ, but the increasing trend survives. 

\begin{table}[htb]
\centering 
\begin{tabular}{c|c|c|c}
 \rule{0pt}{3ex} $z_{\textrm{max}}$ & $ \Omega_{m}$ & $ \beta$ & $\gamma$ \\
\hline 
\rule{0pt}{3ex} \multirow{2}{*}{$0.7$ (109 QSOs)} & $0$ & $6.313$  & $0.680$ \\
\rule{0pt}{3ex}  & $0.305^{+0.379}_{-0.228}$ & $6.472^{+1.533}_{-1.538}$ & $0.674^{+0.049}_{-0.050}$  \\
\hline
\rule{0pt}{3ex} \multirow{2}{*}{$0.8$ (162 QSOs)} & $0.153$ & $5.279$  & $0.714$ \\
\rule{0pt}{3ex}  & $0.511^{+0.305}_{-0.275}$ & $7.162^{+0.715}_{-0.712}$ & $0.651^{+0.023}_{-0.023}$  \\
\hline
\rule{0pt}{3ex} \multirow{2}{*}{$0.9$ (213 QSOs)} & $0.504$ & $5.992$  & $0.690$ \\
\rule{0pt}{3ex}  & $0.548^{+0.295}_{-0.314}$ & $5.912^{+1.155}_{-1.067}$ & $0.692^{+0.034}_{-0.037}$  \\
\hline
\rule{0pt}{3ex} \multirow{2}{*}{$1$ (258 QSOs)} & $0.705$ & $6.331$  & $0.679$ \\
\rule{0pt}{3ex}  & $0.579^{+0.270}_{-0.304}$ & $6.262^{+1.029}_{-1.031}$ & $0.681^{+0.033}_{-0.033}$  \\
\end{tabular}
\caption{Same as Table I of main text, but with the SDSS-4XMM subsample removed.}
\label{QSOvsZ_subsample}
\end{table}

\section{Trend in BAO}
\label{sec:BAO}
Now that we have outlined the increasing $\Omega_m$ trend with redshift, one can ask if it is there in BAO? Note, BAO do not lend themselves so easily to our analytic argument, since BAO represent a statisical statement compressed into an effective redshift. It is possible that any trend gets washed out. Nevertheless, one can quickly confirm a similar feature in BAO below $z=1$. In Table \ref{BAO} we record $D_{M}(z) := c/H_0 \int_0 1/E(z^{\prime}) \dd z^{\prime}$ and $D_{H}(z) := c/H(z)$ constraints from the literature \cite{BOSS:2016wmc, Bautista:2020ahg, Gil-Marin:2020bct, Hou:2020rse, Neveux:2020voa, duMasdesBourboux:2020pck}. Observe that we can divide $D_{M}$ by $D_{H}$ to get 
\be
\label{ratio}
\frac{D_{M}(z)}{D_{H}(z)} = E(z) \int_{0}^z \frac{1}{E(z^{\prime})} \dd z^{\prime}, 
\ee
where $H_0$ and $r_d$ have both dropped out. Note, as we are using anisotropic BAO, it is possible to combine {$D_{M}(z)$ and $D_{H}(z)$ in order to eliminate the $r_d$ factors that customarily appear in constraints}. This leaves a {right hand side in (\ref{ratio}) that depends on $\Omega_m$ only}. It is then easy to solve for $\Omega_m$ and the results are presented in Table \ref{BAO}. To extract the errors, we generate normal distributions for $D_{M}/r_d$ and $D_{H}/r_d$ on the assumption of Gaussian errors, quotient the resulting expressions, {solve for $\Omega_m$, and from the $\Omega_m$ distribution,} identify the $1 \sigma$ confidence intervals. Curiously, BAO beyond $z=1$ agree more closely with the Planck value, $\Omega_m \approx 0.3$. 
%These probes are newer, the BAO peaks are arguably less pronounced and the Lyman-$\alpha$ constraints at $z=2.33$ are still mildly discrepant with Planck-$\Lambda$CDM. As a result, they are regarded as ``non-standard" and are not folded into Planck analysis \cite{Planck:2018vyg}, primarily  on the grounds that the errors are larger. In any case, there seems to be a clash between
%, despite the BAO peak being less well established in these newer observables. 
Furthermore, there is a noticeable clash between 
the Risaliti-Lusso QSOs and BAO beyond $z=1$ in the flat $\Lambda$CDM model. 

\begin{figure}[htb]
   \centering
\includegraphics[width=85mm]{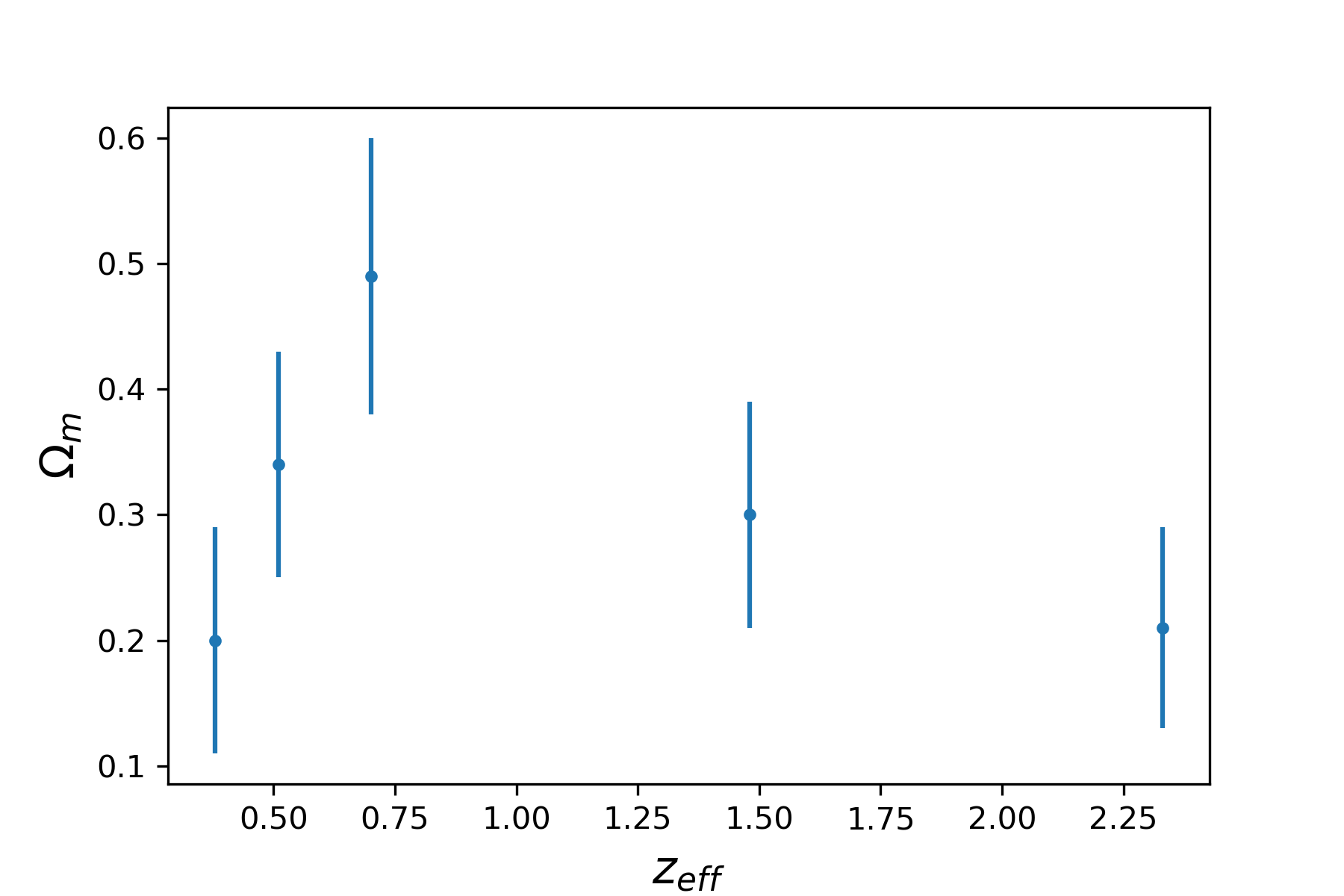}
\caption{Evolution of $\Omega_m$ within existing BAO observations \cite{BOSS:2016wmc, Bautista:2020ahg, Gil-Marin:2020bct, Hou:2020rse, Neveux:2020voa, duMasdesBourboux:2020pck}. 
%The high redshift ($z > 1$) BAO has yet to feature in Planck analysis.
}
\label{OMBAO} 
\end{figure}

It is clear from Fig. \ref{OMBAO} that the trend in the DE dominated regime $z <1$ is pretty pronounced, although it is still not entirely expected. Nevertheless, evidently the value of $\Omega_{m}$ at $z_{\textrm{eff}} = 0.38$ and $z_{\textrm{eff}} = 0.7$ differ by at least $2 \sigma$ and this discrepancy warrants further investigation. If substantiated by DESI \cite{DESI:2016fyo}, this may provide another avenue to confirm the inherent bias in the flat $\Lambda$CDM model, while falsifying the model in the process. 

\begin{table}[htb]
\centering 
\begin{tabular}{c|c|c|c}
 \rule{0pt}{3ex} $z_{\textrm{eff}}$ & $D_{M}/r_d$ & $D_{H}/r_d$  & $\Omega_{m}$ \\
\hline 
\rule{0pt}{3ex}   $0.38$ & $10.23 \pm 0.17$ & $25.00 \pm 0.76$  & $0.20 \pm 0.09$ \\
\rule{0pt}{3ex}   $0.51$ & $13.36 \pm 0.21$ & $22.33 \pm 0.58$  & $0.34 \pm 0.09$ \\
\rule{0pt}{3ex}   $0.70$ & $17.86 \pm 0.33$ & $19.33 \pm 0.53$  & $0.49 \pm 0.11$ \\
\rule{0pt}{3ex}   $1.48$ & $30.69 \pm 0.80$ & $13.26 \pm 0.55$  & $0.30 \pm 0.09$ \\
\rule{0pt}{3ex}   $2.33$ & $37.6 \pm 1.9$ & $8.93 \pm 0.28$  & $0.21 \pm 0.08$ \\
\end{tabular}
\caption{Increasing trend of $\Omega_m$ with effective redshift $z_{\textrm{eff}}$ in BAO results \cite{BOSS:2016wmc, Bautista:2020ahg, Gil-Marin:2020bct, Hou:2020rse, Neveux:2020voa, duMasdesBourboux:2020pck}.}
\label{BAO}
\end{table}

\end{document}